# Deformable Models for Surgical Simulation: A Survey

Jinao Zhang, Yongmin Zhong, and Chengfan Gu

*Abstract*—This paper presents a survey of the state-of-the-art deformable models studied in the literature concerning soft tissue deformable modeling for interactive surgical simulation. It first introduces the challenges of surgical simulation, followed by discussions and analyses on the deformable models, which are classified into three categories: the heuristic modeling methodology, continuum-mechanical methodology, and other methodologies. It also examines linear and nonlinear deformable modeling, model internal forces, and numerical time integrations, together with modeling of soft tissue anisotropy, viscoelasticity, and compressibility. Finally, various issues in the existing deformable models are discussed to outline the remaining challenges of deformable models in surgical simulation.

## I. INTRODUCTION

MODELING and simulation of soft tissue deformation is a fundamental research topic in surgical simulation. Surgical simulation requires realistic and real-time modeling of soft tissue response to tool-tissue interactions [1, 2]; however, it is challenging to satisfy both of these conflicting requirements.

### A. Challenge of Realistic Simulation

The challenge of realistic simulation manifests itself in the context of surgical simulation as the accurate results of material characterization of *in vivo* biological tissues, mesh generation of organ models, and numerical solution to soft tissue behaviors [3]. In order to simulate realistic soft tissue mechanical behaviors, the material properties of living tissues need to be characterized. Such properties are patient-specific, leading to the difficulty in predicting the mechanical behaviors of *in vivo* tissues. Further, the geometry of anatomical models must be acquired from patient-specific medical images, and the problem domain must be discretized. However, this process is still not fully automated, requiring significant labours in image segmentation and mesh generation. Finally, numerical issues for soft tissue simulation must be overcome. Most biological soft tissues exhibit anisotropic and heterogeneous stiffness and are nearly incompressible [4], leading to ill-conditioned problems during simulation. Such problems are numerically expensive to solve and involve inaccuracy or even instability in simulation. Boundary conditions are also difficult to define due to complex tissue compositions and interactions between organs.

### B. Challenge of Real-Time Simulation

A simulation that is mechanically realistic but not interactive would not fit for surgical simulation. The challenge of real-time simulation manifests itself as the high solution speed to tool-tissue interactions. Soft tissue response must be computed in a short time to achieve the required update rates of visual and haptic feedback for real-time user interaction with virtual tissue models. The update rate required for visual feedback is at least 30 Hz to achieve continuous motion of rendered graphics to human sensory system, whereas the update rate required for haptic feedback is at least 1,000 Hz to achieve stable and smooth tactile rendering from the haptic device [5]. Due to nonlinear characteristics of soft biological tissues, the numerical solutions are often computationally expensive to obtain. The need for real-time computation, however, often requires simplifications of the problem, adversely affecting the simulation accuracy [6].

In all, the realistic and real-time characteristics of surgical simulation not only pose challenges to each aspect but also affect each other mutually since performance improvement on one aspect is mainly achieved by the detriment of the other.

This paper presents a survey of the current state-of-the-art deformable models for modeling and simulation of soft tissue deformation in interactive surgical simulation. The purpose of this survey is not only to present a review on deformable models used in surgical simulation but also to reflect the recent progress in the field of soft tissue modeling since the precedent surveys [7-12]. The scope of this survey is focused on soft tissue response induced by mechanical event and associated deformation only, which occurr commonly in tool-tissue interactions such as pushing, prodding, and palpation. Although soft tissue deformation induced by other events, such as thermal and electrical events, is beyond the scope of this survey, they can be simulated in conjunction with mechanical event at the expense of added computational complexity. Since surgical simulation is a research area involving multiple research topics, this paper focuses on deformable models for surgical simulation. Readers can find detailed information on other related topics such as continuum biomechanics of soft biological tissues [13], haptic feedback [14-17], augmented reality in surgery [18], physically-based simulation of cutting [19], collision detection [20], and computational biomechanics

J. Zhang is with the School of Engineering, RMIT University, Bundoora, VIC 3083, Australia (e-mail: jinao.zhang@rmit.edu.au).
Y. Zhong is with the School of Engineering, RMIT University, Bundoora, VIC 3083, Australia (e-mail: yongmin.zhong@rmit.edu.au).
C. Gu is with the School of Engineering, RMIT University, Bundoora, VIC 3083, Australia (e-mail: chengfan.gu@rmit.edu.au).

model generation [21].

## II. Deformable Models for Surgical Simulation

This section presents a survey of the deformable models developed for interactive surgical simulation. It divides the deformable models into three basic categories: the heuristic modeling methodology, continuum-mechanical methodology, and others. The first category is made up of heuristic models that are derived from rather straightforward modeling schemes for the geometry of soft tissues, allowing for the inclusion of elastic properties. The second category contains deformable models that account for the deformation of soft tissues from the viewpoint of continuum mechanics and describe the mathematical terms by equations of solid mechanics. According to the use and nonuse of mesh, it is further divided into the mesh-based approach and meshless approach. Finally, the third category consists of deformable models that are based on other concepts for soft tissue deformation, such as the neural network method, machine learning, data-driven approach, and fibers-fluid technique.

### A. Heuristic Modeling Methodology

#### 1) Geometrically-based Models

In the early efforts on modeling of soft tissue deformation for surgical simulation, various geometrically-based approaches, such as the free-form deformation (FFD) [22] and deformable splines [23], were studied owing to their computational advantages. In 1986, Sederberg and Parry [22] proposed a lattice-based FFD technique to deform a soft tissue surface model via a parametric parallelepiped lattice based on the manipulation of control points in a free-form manner (see Fig. 1(a)). As the position of control points changes, the surface of the free-form lattice is deformed using the tensor product of a tri-variate Bernstein polynomial to determine the displacement of points on the lattice surface. Global and local deformations can be obtained through the manipulation of control points. Cover et al. [23] studied a technique of deformable splines for soft tissue deformation and further applied it to simulate laparoscopic gall-bladder surgery. This technique induces deformation on a soft tissue surface model by minimizing the potential energy, which is proportional to the degree of elastic deformation, with respect to the displacement of control points to achieve the corresponding deformation state (see Fig. 1(b)). Although the geometrically-based models are often fast for interactive soft tissue deformation, they do not provide a realistic simulation of soft tissue mechanical behaviors to meet the accuracy requirement of surgical simulation, since the deformation is carried out indirectly via the manipulation of control points with no resemblance to the physical behaviors underlying soft tissue deformation. As such, the geometrically-based models have been mainly superseded by physically-based deformable models, which consider physical properties of materials and physical dynamics to improve simulation accuracy and to obtain a satisfactory degree of physical realism.

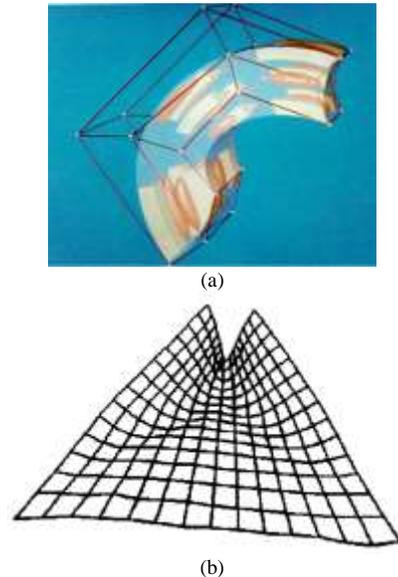

(a)

(b)

Fig. 1. (a) Deformed state of an FFD lattice with control points [22]; and (b) deformed surface using deformable splines [23].

#### 2) Mass-Spring Model

Mass-spring model (MSM) [24, 25] is a popular deformable model based on the principle of dynamics for computation of soft tissue deformation [8, 9, 26, 27]. It is widely used for modeling of soft tissue deformation to simulate various kinds of surgical procedures such as the repair of heart valves [25], transurethral resection of prostate [28], and endoscopic surgery [29]. As illustrated in Fig. 2, MSM considers a soft tissue model as a network of lumped masses connected via elastic springs. The dynamics of soft tissue deformation are governed by the non-rigid mechanics of motion, in which the internal force at a mass point is due to the sum of spring forces via elastic springs connected to this point. The positions of mass points are obtained by considering the balance of force through time-stepping in temporal domain. MSM is simple in implementation and efficient in computation, leading to an effective means for modeling of soft tissue deformation for interactive surgical simulation. Soft tissue mechanical properties, such as heterogeneity, near incompressibility, and time-dependent viscoelasticity, can be realized by techniques such as the modification of spring stiffness constants [30, 31], utilization of penalty forces [32], and incorporation of mechanical dampers [33], respectively. The literature on soft tissue deformation using MSM is abundant, and various improvements were proposed to enhance the capabilities of MSM. Compared to the conventional MSM where elastic springs are governed by the linear Hooke's law, Basafa and Farahmand [34] employed a piecewise nonlinear spring model with a two-step expression of force-displacement relationship for modeling of nonlinear soft tissue deformation in laparoscopic surgery. The spring force is formulated by considering the typical soft tissue nonlinear force-displacement relationship made by a "toe" region at small deformation and a region of constant stiffness at large deformation. Qin et al. [35] improved MSM by constructing a multi-layered MSM based on the layered structure of biological soft tissues and further applied this technique in the virtual orthopaedic surgery. Choi et al. [33, 36, 37] devised a force propagation MSM for virtual reality (VR) based medical learning. It considers the process of soft tissue deformation as a process of force propagation among the masses of soft tissues on a per-node basis [36]. A penetration depth is employed to limit the range of force propagation for the benefit of computational efficiency. However, this penetration depth is determined subjectively, relying on detection of unnoticeable change in shape. Further, the determination of the penetration depth does not consider material properties. Omar et al. [38] reported a local deformation method based on MSM. Based on elastic theory, this method estimates the stress distribution in soft tissues according to a depth from the contact surface subject to an external load. Subsequently, the local deformation range, to which MSM is applied, is determined from this stress distribution. Duan et al. [24] applied deformable constraints to MSM to directly manipulate the position of mass points to satisfy a set of predefined geometric and volume constraints for nonlinear force-displacement characteristics and near incompressibility of soft tissues. Omar et al. [39] reported a nonlinear MSM using conical springs to replace linear Hookean springs for soft tissue deformation; however, the use of the conical spring increases the computational load and also involves more mechanical parameters.

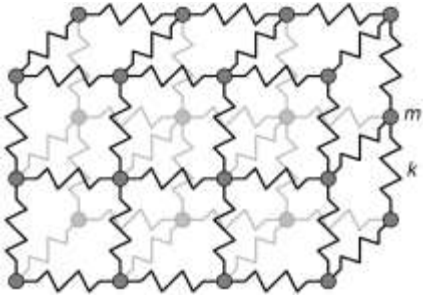

Fig. 2. A portion of a MSM: points of lumped mass $m$ is connected via a network of elastic springs of stiffness $k$ [27].

Despite improved physical realism offered by the principle of dynamics, MSM suffers from a number of deficiencies that limit its model accuracy. In general, mass-spring systems are not convergent as the mesh is refined, meaning the solution of deformation does not converge on the true solution [26]. Instead, the geometrical structure and topological arrangement of elastic springs heavily influence the deformation behaviors of the model and may introduce artificial anisotropy and heterogeneity [40], giving rise to stability and accuracy issues [3]. Further, regardless of linear Hookean or nonlinear springs, the internal nodal force is determined by the sum of the spring forces, which are dependent only on the position of neighboring points, spring rest lengths and spring stiffness constants; therefore, the mechanical behaviors of individual springs cannot be simply related to the constitutive laws governing the mechanical behaviors of soft tissues. Consequently, the nonlinear stress-strain relationship of soft biological tissues is difficult to be reproduced accurately by MSM. Owing to this, optimization algorithms such as the simulated annealing (SA) [41] and Genetic algorithms (GA) [42] are often employed for optimization of spring stiffness constants by fitting the deformation of MSM to some reference data to achieve certain global mechanical behaviors. However, parameter optimization is a tedious task, and the result of a particular optimization may no longer be valid if model topology arrangement and boundary conditions are changed. Overall, the popularity of MSM in surgical simulation is mainly attributed to its simple mesh structure, easy programmability, and low computational complexity; as the evolution of MSM already reached its peak [8], it is expected that the application of MSM in surgical simulation will be superseded by other deformable models that have higher physical realism with real-time computational efficiency.

### 3) ChainMail Algorithm

Compared to MSM, ChainMail algorithm [43, 44] is a more simplified approach for soft tissue deformation. In the early years of computer graphics, objects were commonly represented by surface-based polygonal models; despite their computational efficiency, these surface models are less accurate than modeling of object volume for soft tissue deformation [6]. ChainMail algorithm was proposed under this background, which considers the volumetric nature of human organs with a deformation law derived from MSM, forming a linked volume to describe volumetric behaviors of soft tissues [44]. The basic unit in the ChainMail algorithm is called the chain element, which occupies the position of a voxel in a linked volume model. Each chain element enforces a geometric bounding region formed by geometric limits to each of its neighboring chain elements. The position of a chain element will be adjusted to satisfy geometric constraints only if the position is outside of the bounding region enforced by its neighbors (see Fig. 3). This position adjustment mechanism is further followed by a relaxation scheme that minimizes the global potential energy of the system. The ChainMail algorithm can simulate various soft tissue mechanical behaviors, such as the nonlinear force-displacement relationship, hysteresis, and stress relaxation [45]. One significant advantage of the ChainMail algorithm is its computational efficiency afforded by the position adjustment mechanism, and hence it has been used extensively in modeling of large medical volume deformation consisting of millions of voxels, each of which stores important information related to patient-specific tissues and organs [46-48]. Such large volumetric datasets cannot be interactively deformed by conventional deformable models. Even with significant mesh processing, the computational complexity of these conventional approaches limits the resolution of the captured medical datasets to only a small fraction that is usually several orders of magnitude lower, resulting in an inevitable loss of details of source data [46].

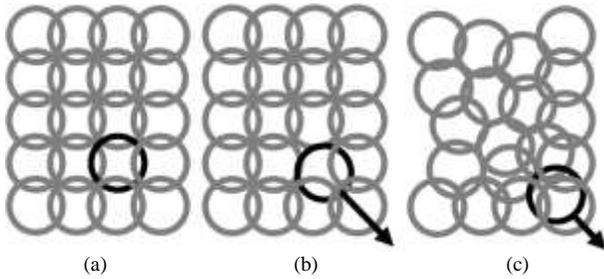

Fig. 3. Deformation of the ChainMail algorithm: (a) a chain element (black) at undeformed state; (b) the chain element is moved along the path of the arrow; and (c) its neighboring chain elements move to satisfy geometric constraints between elements, leading to a deformed state.

Since the inception of the ChainMail algorithm, extensive research efforts have been dedicated to the improvement of its physical accuracy and computational performance. Schill et al. [49] presented an enhanced ChainMail to simulate the vitreous humor in the eye [50], a substance that is heterogeneous and highly deformable. This enhanced ChainMail extends the traditional ChainMail to modeling of heterogeneous materials. Park et al. [51] proposed a shape-retaining 3D ChainMail, or S-Chain in short, for real-time haptic rendering. The haptic force is calculated based on the idea that the reflection force is proportional to the sum of the distances of all moved chain elements. To address the issue of geometric degradation due to shear distance limit in the traditional ChainMail, Wang and Fenster [52] studied a restricted 3D ChainMail, which replaces the shear distance limit by an angular shear limit expressed in degree, confining the movement of a chain element within a ChainMail bounding trapezium or frustum in 2D or 3D, respectively. Li et al. [53] proposed a surface ChainMail for web-based surgical simulation, which enhances the traditional ChainMail by defining the stretching, compressing and shear limits using a strain limit in relation to the rest length between two chain elements. Further, Li and Brodlie [54] devised a generalized ChainMail that can be applied to any type of grid. In this method, the chain elements can be arbitrarily positioned and linked to any number of neighboring chain elements, extending the range of applications of the traditional ChainMail algorithm. To achieve a more accurate deformation with physical meanings, Wang and Lu [55] presented an adaptive S-Chain, utilizing an energy-based wave propagation on the object surface, whereas the inner volume is deformed by the S-Chain. Based on the generalized ChainMail, Levin et al. [56] proposed a ChainMail-mass-spring hybrid model, where the ChainMail constraints are employed for checking constraint violations and spring forces are calculated once the ChainMail constraints are satisfied. Duysak and Zhang [57] studied a mass-spring chain model, combining the strengths of both MSM and ChainMail algorithm. This model applies the ChainMail constraints to a triangular surface mesh, confining the movement of a spring within a ChainMail bounding region made by super elastic limit, rigid limit, minimum spring length and maximum spring length. Neubauer [58] studied a ChainMail algorithm, which is named the Divod ChainMail, for direct volume deformation. Rodriguez et al. [48] proposed an

SP-ChainMail which implements the ChainMail algorithm on the Graphics Processing Unit (GPU). This method can achieve a speed gain of more than 20x when using a modern GPU compared to that of the Central Processing Unit (CPU) counterpart. The SP-ChainMail is further extended by the heterogeneous SP-ChainMail [46] to simulate heterogeneous materials and handle multiple concurrent deformations. Zhang et al. [59] presented a time-saving volume-energy conserved ChainMail, which conserves both volume and strain energy for soft tissue deformation. This method also improves the computational time for isotropic and homogeneous materials, since it considers each chain element only once for position adjustment via a time-saving scheme.

Thanks to various improvements, the ChainMail algorithm has been applied to many medical applications such as the arthroscopic knee surgery [60], intra-ocular surgery [61], web-based surgical simulations [53, 54], prostate brachytherapy simulation [62], virtual endoscopy applications [58], training simulators with respiratory components [63], angioplasty simulation [64], percutaneous transhepatic cholangio-drainage (PTCD) simulation [47] and image-based palpation simulation [65]. Despite various applications in the field of surgical simulation, the ChainMail algorithm suffers from the empirical selection of parameters for geometric constraints [59]. Further, since the ChainMail relies solely on element positions rather than the equations of motion to determine element displacements, the dynamic behaviors of soft tissues are difficult to realize.

### 4) Others

Other deformable modeling methods such as the shape matching technique coupled with position-based solver [66] were also studied for soft tissue deformation. Shape matching is a geometrically-motivated approach based on finding the least squares optimal rigid transformations between two sets of points with prior knowledge of correspondence [67]. Similar to the ChainMail algorithm, the shape matching technique also directly manipulates the position of points to satisfy a set of geometric constraints. However, this method relies on determination of an optimal cluster stiffness coefficient for realistic soft tissue deformation, without considering the material properties of soft tissues.

To sum up, the deformable models in this category share a common characteristic, that is, they suffer from ambiguity in specification of appropriate model parameters to reproduce the mechanical behaviors of soft tissues due to discrete nature of the model, leading to unclear and not well-defined relationships between model parameters and material constitutive laws. An optimization process is often required to tune model parameters [68] in order to achieve a certain level of physical accuracy. Since this optimization process relies on a reference solution of defined model topology and boundary conditions, the resultant model parameters may become invalid if one of these conditions is changed during the simulation. Further, as mentioned in Section II.A.2, these modeling methods are not convergent as the mesh is refined, posing significant physical accuracy issues in comparison with the continuum-mechanical methodology to be introduced next.

### B. Continuum-Mechanical Methodology

Different from the heuristic modeling methodology which assumes a discrete model representation of soft tissues for deformable modeling, the continuum-mechanical methodology considers soft tissues as a continuum medium based on the continuum mechanics of solid and employs constitutive laws to account for the complex mechanical behaviors of soft tissues. The solution procedure of this methodology typically involves the consideration of minimization of overall potential energy and/or other fundamental physical balance laws to determine unknown field variables over the problem domain. The typical solution methods can be further divided into two sub-categories based on the use and non-use of object mesh, such as the finite element method (FEM) and boundary element method (BEM) in the category of mesh-based approach, and meshless total Lagrangian explicit dynamics (MTLED) algorithm and smoothed particle hydrodynamics (SPH) in the category of meshless approach.

#### 1) Mesh-based Approach

##### 1.1) Finite Element Method

FEM is a typical method for simulation and analysis of soft tissue deformation in surgical simulation, which requires explicit construction of the object mesh to approximate the constitutive laws governing the mechanical behaviors of soft tissues. In FEM, an approximate discrete representation of the target soft tissue can be obtained by dividing the soft tissue model into a number of elementary building components called the finite elements, forming a finite element mesh of triangular or quadrilateral elements in 2D or tetrahedral or hexahedral elements in 3D that conforms to the problem domain (see Fig. 4). The constitutive laws are approximated with respect to each finite element and satisfied at element level. The individual equations of the finite elements under external loads are assembled into a large system of equations that represents the mechanical behaviors of the entire soft tissue model, from which the nodal displacements are subsequently determined [69]. Soft tissue material properties, such as the Young's modulus and Poisson's ratio, can be obtained by experimental measurements and directly integrated into the parametric constitutive laws for finite element calculation. Due to its physical accuracy, FEM is popular in the computational biomechanics [70] and has been applied successfully into a wide range of biomechanical modeling of soft tissues, such as the modeling of soft tissue deformation in image-guided hepatic surgery [71], computer-integrated neurosurgery [72], whole-body medical image registration [73], and interventional electrocardiology procedures [74]. Although FEM can achieve high model accuracy, such accuracy is obtained at the expense of high computational cost, leading to great challenges for interactive soft tissue deformation.

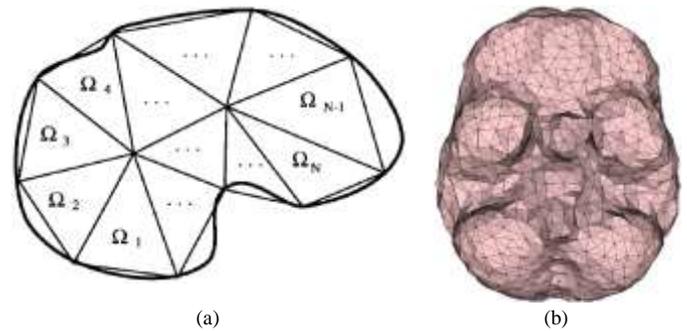

(a)            (b)

Fig. 4. (a) The problem domain is divided into a number of triangular finite elements $\Omega_N$ [75]; and (b) a brain model is discretized into a tetrahedral mesh [76].

##### 1.2) Simplification of FEM

To meet the real-time computational performance for surgical simulation, various techniques were proposed to simplify the computational complexity of FEM. The explicit FEM [26] is a simplified form of FEM and is often employed for soft tissue deformation. It can be integrated either explicitly or implicitly [26]. In the explicit FEM, the internal and external forces and masses are lumped to the nodes, leading to block diagonal mass and damping matrices through which the computation can be performed at element level [77], allowing simple implementation and easy parallel computation. Bro-Nielsen and Cotin [78] developed a fast finite element (FFE) model which simplifies the computational complexity of FEM using matrix condensation [75]. By condensing the full system matrix describing the behaviors of object volume to a new matrix that involves only the variables of surface nodes while preserving original physical characteristics of the volumetric model, the computational time for the deformation of volumetric model can be reduced to the computational time of a model only involving surface nodes of the mesh, leaving only the displacements of the boundary nodes as unknowns [79]. However, despite the improved computational efficiency, this simplification significantly degrades the simulation accuracy. Cotin et al. [80] applied a pre-computation technique to the linear FEM to achieve real-time computational performance for hepatic surgery simulation; the equilibrium solutions of a linear FEM model are pre-computed, and the principle of superposition is applied to determine nodal positions at interactive frame rates. However, the pre-computed elementary deformations are only valid for a given configuration of the stiffness matrix. Cotin et al. [81] also proposed a tensor-mass model (TMM) which incorporates the concept of shape function in FEM into the formulation of internal force and obtains nodal positions in a mass-spring fashion. This model simplifies the computational complexity of FEM to that of MSM while retaining the calculation of internal force to be independent of mesh topology. BEM [82, 83] simplifies the FEM complexity by formulating the weak form of the principle of virtual work into a surface integral form based on the assumption of an isotropic and homogeneous material interior. Owing to this assumption, the deformation solution is reduced to that of the boundary integration equation on surface mesh only, which significantly facilitates the computational performance. However, BEM only works for objects whose interior is composed of isotropic and homogeneous materials [8], and hence it cannot accommodate the anisotropic and heterogeneous characteristics of soft tissues. Wang et al. [84] applied BEM into a surgical simulation for haptic deformation of soft tissues and surgical cutting. Zhu and Gu [83] applied BEM into a mass-spring constraint model, where BEM is used to determine the global deformation and MSM to interactively simulate the dynamic behaviors of soft tissues. Inspired by the concept of geometric constraints used in the heuristic modeling methodology, Tang and Wan [85] studied a strain-limiting FEM for virtual surgical training. This model reduces the FEM complexity from solving a system of equations to solving a set of geometric constraints by using a series of strain-limit constraints on the principle strains of the strain tensor. A multi-resolution hierarchy mesh structure is also employed to facilitate the global convergence of the constrained system. Despite the computational advantage, the utilization of strain limits in this method adversely confines the deformation range of the finite elements. Liu et al. [86] coupled FEM computation of strain energy density function with MSM internal force calculation for modeling of soft tissue deformation. In this method, the spring force is calculated based on the strain energy density at neighboring points, which is determined from the strain energy of the tetrahedron in the finite element mesh. Soft tissue mechanical behaviors can be predicted by various forms of strain energy density function. Goulette and Chen [87] presented a hyperelastic mass links (HEML) algorithm for fast computation of soft tissue deformation. HEML is derived from the framework of FEM but calculates nodal positions via a mass-spring fashion based on local nodal displacements. It obtains a speed gain more than 40x compared to TMM. Zhang et al. [88] proposed an energy balance method (EBM) based on the law of conservation of energy for modeling of soft tissue deformation. In this method, the work-energy balance is achieved via a position-based incremental process for the new equilibrium state of soft tissues. EBM employs nonlinear geometric and material formulations to account for nonlinear soft tissue deformation. It can accommodate anisotropy, viscoelasticity and near incompressibility of soft tissues via various strain energy density functions.

### 1.3) Total Lagrangian Formulation

Considering the frame of reference, FEM employs two formulations which are the updated Lagrangian formulation and total Lagrangian formulation for determination of the unknown values of state variables [89]. In the updated Lagrangian formulation, all variables are referred to the current system configuration from the end of the previous time step. The advantage of this formulation is the simplicity of incremental strain description and low internal memory requirements [72]; however, it requires a re-calculation process of spatial derivatives at each time step, since the reference configuration varies with time. This re-calculation process is computationally expensive, unsuitable for real-time computational performance of surgical simulation. Compared to the updated Lagrangian formulation, the total Lagrangian formulation considers all variables referred to the initial system configuration. Contrary to the incremental strain description, the strain formation in the total Lagrangian formulation leads to correct results after a load cycle, without occurrence of error accumulation [90]. More importantly, it enables all derivatives with respect to spatial coordinates to be pre-calculated and stored [91], since the initial configuration is explicitly defined and does not change with time. It takes 10.6 ms for the updated Lagrangian framework to find a solution for meshes with 2,535 nodes and 2,200 hexahedrons under ellipsoid indentation, while only 2.1 ms for the total Lagrangian framework [92]. Based on the computational advantage, Miller et al. [77] developed a total Lagrangian explicit dynamics (TLED) finite element algorithm for soft tissue deformation, achieving fast solution calculation through the pre-computation of spatial derivatives, element-level computation, and explicit time integration. Given these three important attributes of TLED, it can be easily parallelized on GPU to take advantage of hardware parallel computation. Taylor et al. [91] reported a method to achieve a high-speed TLED solution with GPU parallel computing. This method can obtain a speed gain up to 16.8x compared to the equivalent CPU implementation. Later, Taylor et al. [93] achieved a speed gain of 56.3x using NVIDIA Compute Unified Device Architecture (CUDA) implementation, which is subsequently integrated into the GPU-based finite element package NiftySim [94]. The GPU-accelerated TLED has been successfully applied into the simulation of neurosurgical procedures [89, 95], whole-body computed tomography (CT) image registration [73] and non-rigid neuroimage registration [96]. Its computational potential has also been analyzed utilizing a wide range of GPUs [97]. Szekely et al. [90] applied the total Lagrangian formulation-based FEM into the simulation of uterus deformation. Based on the total Lagrangian formulation, Marchesseau et al. [98] also presented a multiplicative Jacobian energy decomposition (MJED) approach to discretizing hyperelastic materials on linear tetrahedral meshes. This approach decouples in the strain energy the invariants of the right Cauchy-Green deformation tensor from the Jacobian so as to avoid matrix inversion and complex derivative expressions, leading to faster matrix assembly than the standard FEM. However, MJED requires a decomposition of the strain energy into simple terms such that pre-computation can be performed to speed-up the assembly of stiffness matrices. Mafi and Sirouspour [99] also developed a total Lagrangian formulation-based FEM algorithm coupled with GPU-based implicit dynamics for soft tissue deformation. This GPU-based solution addresses the real-time computational challenge in both areas of FEM matrix construction and solving the system matrix resultant from the implicit integration.

### 1.4) Model Reduction

In addition, model reduction techniques [100, 101] have also been applied to FEM for achieving improved computational efficiency. The essential idea is to employ a set of global basis, that is, in a statistical sense, the best suited to reproduce the complete models, by which the full system response is projected into a smaller dimensional subspace, leading to a reduction in the number of degrees of freedom for deformation calculation. This is in sharp contrast with the standard FEM, which employs piecewise polynomial shape functions to approximate the solutions in the Galerkin framework [102]. The solution procedure of model reduction is made up by two steps: one is the offline step, in which the response of soft tissues to prescribed loads is extracted to construct a meta-model and stored in the memory; the other is the online step, in which the model is interpolated for any other load state to perform a reduced-model simulation with smaller degrees of freedom. Niroomandi et al. [103] studied a model reduction technique based on the proper orthogonal decompositions (POD) for simulation of palpation of human cornea with surgical tools. Despite the improved computational efficiency, the reduced model in this approach is actually linear since no update of the tangent stiffness matrix is performed, resulting in higher strains in comparison with the standard FEM model. Owing to this deficiency, Niroomandi et al. [104] coupled the POD with a nonlinear solver, the asymptotic numerical method (ANM), to construct a geometrically nonlinear reduced-order model for soft tissue deformation. Later, Niroomandi et al. [105] generalized their POD approaches by considering a parametric problem using the proper generalized decomposition (PGD) to simulate liver deformation under interaction with surgical scalpels. As a generalization of PODs, the resulting PGD solution is expressed as a finite sum of separable functions that provides a meta-model to real-time obtain the response of the system at kilohertz rates. Radermacher and Reese [106] also facilitated the computational performance of POD by using the discrete empirical interpolation method (DEIM), which works as an additional treatment to further reduce the nonlinear terms based on a small number of interpolation indices. This model can obtain a speed gain of 10x compared to the classical POD methods without empirical interpolation. A review of model reduction techniques for computation of soft tissue deformation can be found in [107].

### 1.5) Element-Related Issues

When applying the mesh-based approach to compute soft tissue deformation, it is important to consider the element-related issues to avoid numerical deficiencies. In order to satisfy the computational requirement of surgical simulation, the finite element models must use numerically efficient low-order elements, such as the eight-node linear under-integrated hexahedrons and four-node linear tetrahedrons [95]. However, it is known that the standard formulation of these elements exhibits numerical deficiencies. The eight-node linear under-integrated hexahedrons exhibit the zero energy mode, where individual elements are deformed while the overall mesh is undeformed, resulting in hourglass-like element shapes. Joldes et al. [92] proposed an hourglass control algorithm based on the total Lagrangian formulation to eliminate the zero energy mode (see Fig. 5). The standard formulation of the four-node linear tetrahedrons exhibits artificial stiffening when simulating nearly incompressible materials such as biological soft tissues, referred to as volumetric locking [94]. Joldes et al. [108] addressed this issue by using an improved average nodal pressure (IANP) linear tetrahedron formulation (see Fig. 6).

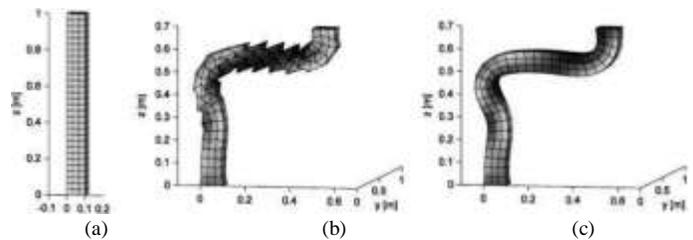

Fig. 5. Hourglass control in a deformed column: (a) undeformed shape; (b) deformed shape without hourglass control; and (c) deformed shape with hourglass control [92].

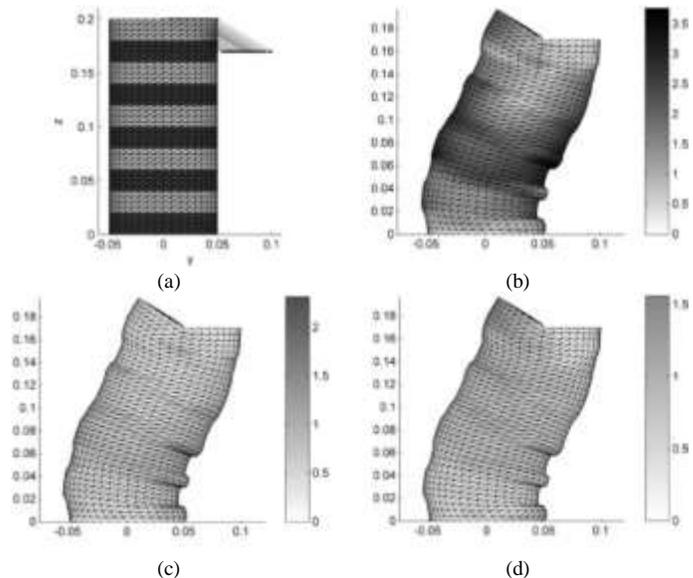

Fig. 6. Volumetric locking control in a deformed cylinder: (a) undeformed shape with prescribed nodal displacements; (b) locking tetrahedral elements; (c) average nodal pressure (ANP) elements; and (d) IANP elements; the color bars show the position difference of surface nodes to the reference solution using hexahedral elements [108].

### 2) Meshless Approach

#### 2.1) Shortcomings of the Mesh-based Approach

Despite the popularity and high level of accuracy of finite element-based methods in computation of soft tissue deformation for surgical simulation, the result of these methods heavily relies on the quality of the object mesh that discretizes the model geometry. In consideration of physical accuracy and numerical convergence, a good quality mesh is always required. However, owing to the complex geometry of a human organ, it is very difficult to build a good quality mesh automatically. It commonly requires an experienced analyst to manually create a quality mesh, leading to a labor-consuming problem. The literature [109] shows that it takes more than two months for an experienced analyst to create a good hexahedral mesh, which is a major bottleneck in the efficient generation of patient-specific models used for real-time simulation of surgical procedures. Even if a good quality mesh is generated, the solution method may still fail in the case of large deformation where elements become highly distorted during the loading process, resulting in element inversion [110] with zero or negative Jacobians [5].

## 2.2) Meshless Total Lagrangian Explicit Dynamics

Compared with the mesh-based approach, the meshless approach [109, 111-113] conducts object deformation without involving the mesh topology of the discretized soft tissue model, overcoming the degradation of mesh quality at large deformation involved in the mesh-based approach [114]. It uses a set of particles (mass points) dispersed arbitrarily in the problem domain and interpolates the state variables of each particle through consideration of state variables at neighboring particles (see Fig. 7). Based on the total Lagrangian formulation where pre-computation can be performed, Horton et al. [109] proposed a meshless total Lagrangian explicit dynamics (MTLED) algorithm in the element-free Galerkin (EFG) framework. Numerical integration is conducted through the theory of moving least-squares (MLS) with the aid of hexahedral background integration cells that are not conformed to the simulation geometry. As in the TLED finite element algorithm, the MTLED applies pre-computation of all derivatives with respect to spatial coordinates of each integration cell and uses the deformation gradient to determine the full system matrix at each time step. With the same number of nodes, the presented MTLED runs at half the speed of a hexahedral-based TLED simulation but three times faster than a similar tetrahedral-based simulation [109]. However, the standard meshless shape functions are generally not polynomials. They are created on overlapping support domains, which are constructed using support nodes located beyond the boundary of integration cells. In many cases they are not interpolatable at nodes, leading to significant challenges in numerical integrations and enforcing boundary conditions. Further, the use of hexahedral background integration cells in the MTLED may induce volume inaccuracy when the hexahedral cells are intersected by a domain boundary due to the complex geometry of the human organ. Zhang et al. [115] addressed the issue of volume inaccuracy by employing tetrahedral background integration cells to improve the accuracy of volumetric integration. The MLS is also coupled with the finite element shape functions to impose essential boundary conditions. However, blending the meshless approach with the finite element shape functions requires the creation of a finite element layer along the essential boundary, which is actually one of the issues that the meshless approach aims to avoid. Joldes et al. [116] eliminated this issue by applying displacement corrections in a prediction-correction manner to enforce essential boundary conditions. Their approach is especially efficient in the total Lagrangian framework since the parameters related to the determination of computed corrections can be pre-computed. Zou et al. [117] employed a radial point interpolation method (RPIM) for easy enforcement of essential boundary conditions, since the shape functions of the RPIM have the Kronecker delta function property. This method is further applied to a neurosurgical simulation. To further improve the numerical accuracy of MTLED, Chowdhury et al. [118] studied a modified MLS (MMLS) algorithm which uses a second-order polynomial basis to generate more accurate approximation of deformation fields for randomly distributed nodes. For the same size of supporting domains, MMLS can generate more accurate results than the classical MLS with linear basis, and it is more efficient in computation since the radius of influence does not need to be as large as that in the classical MLS with quadratic basis. Despite the advantage of MTLED in handling large deformation of soft tissues, MTLED should not be used when the reaction force and displacement of a single node are needed, since this method is accurate in terms of the overall reaction forces but not quite good with individual displacements or forces [109].

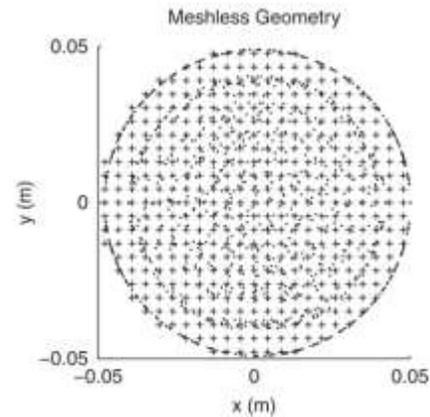

Fig. 7. A meshless geometry: nodes (·) are dispersed arbitrarily in the problem domain with background integration points (+) [109].

### 2.3) Smooth Particle Hydrodynamics and Point-Collocation-based Method of Finite Spheres

Different from MTLED which uses a grid of background cells for numerical integration, particle-based methods such as the smoothed particle hydrodynamics (SPH) and point-collocation-based method of finite spheres (PCMFS) were also studied for soft tissue deformation. In these methods, the particles in the problem domain have an associated smoothing distance over which the state variables are interpolated by a kernel function (see Fig. 8) in consideration of state variables at neighboring particles. Palyanov et al. [119] presented a predictive-corrective incompressible SPH (PCISPH) algorithm (named Sibernetic) for biological soft tissue simulation, whereas Rausch et al. [120] employed a normalized total Lagrangian SPH, taking the natural ability of meshless approach in creating material discontinuities for modeling of soft tissue damage and failure. Similar to SPH, the discrete element method (DEM) [121, 122] and Peridynamics [123, 124] also provide a good alternative to SPH for soft tissue deformation, since the motion of a particle is also computed by summing the forces of neighboring particles. DEM uses a discrete representation where the particles are modeled as individual entities, and it requires calibration of microscopic parameters. Peridynamics formulates constitutive models in continuum mechanics as the relationships between deformation state and force state, and it can model soft materials with any Poisson's ratio.

However, the related research on using these two methods for soft tissue deformation has been very limited. De et al. [40] applied a localized linear PCMFS model for real-time soft tissue deformation based on the assumption that the surgical tool-tissue interaction is local and the deformation field dies off rapidly with increase in distance from the surgical tool tip. Later, Lim and De [5] extended the localized linear PCMFS by considering geometric nonlinearity to enhance the response of the linear model with nonlinear deformation in the local neighborhood of surgical tool-tip. To further extend the PCMFS to accommodate nonlinear characteristics of soft tissues, Banihani et al. [125] applied the POD technique to the PCMFS with consideration of hyperelastic materials.

Despite the advantage of meshless approach in handling large deformation and discontinuities, it generally has difficulty in handling sparely sampled regions [21], and its physical accuracy heavily relies on the proper placement of sampled nodes [40, 109].

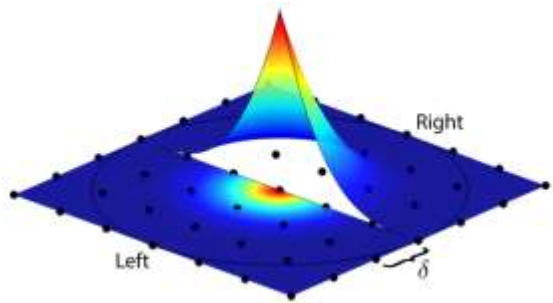

Fig. 8. A kernel function: (right) the kernel's weight distribution; and (left) projection of the kernel function onto the 2D plane to illustrate the radius of influence (in this case three times the particle distance δ) [120].

### C. Other Modeling Methodologies

Apart from the modeling methods mentioned above for soft tissue deformation for surgical simulation, a few other methods based on different concepts were also studied for soft tissue deformation. Zhong et al. [126] proposed a cellular neural approach by taking the real-time computational advantage of cellular neural network for interactive soft tissue deformation. The soft tissue deformation is carried out from the viewpoint of potential energy propagation [127], in which the mechanical load of an external force applied to soft tissues is considered as the equivalent potential energy, according to the law of conservation of energy, and it is propagated in the soft tissue through CNN-based neural propagation [128]. De et al. [129] also studied a neural network technique of machine learning for soft tissue deformation. The computational process is divided into an offline and an online phase, in which the offline phase pre-computes the response of a FEM model subject to prescribed displacements and optimizes the coefficient of neurons through training of a radial basis function network (RBFN); and the online phase reconstructs the deformation field using the trained RBFN. The concept of machine learning for computation of soft tissue deformation is further explored by Lorente et al. [130] for modeling of liver deformation during breathing. In this approach, deformation data are used to feed a supervised machine learning model to find a mapping function of input variables to approximate known outputs. This mapping function is constructed and capable of generating an output, i.e., the global deformation of soft tissues, for future unseen inputs, i.e., the prescribed displacements. Therefore, the performance of the machine learning model is highly dependent on the training data and chosen learning algorithm. Tonutti et al. [131] further studied the capability of machine learning in real-time modeling of soft tissue deformation, showing that the use of machine learning for soft tissue deformation can achieve position errors below 0.3 mm which is beyond the general threshold of surgical accuracy in image-guided neurosurgery. Similarly, Bickel et al. [132] studied a data-driven approach based on a linear co-rotational FEM for soft tissue deformation. The deformations of a real object are captured and each of the captured deformations is presented as a spatially varying stress-strain relationship in the FEM model. Material properties are then interpolated from this stress-strain relationship in the strain-space. The data-driven approach also provides a feasible solution to the problem of model parameter identification mentioned previously, leading to the models with increased accuracy to describe soft tissue deformation behaviors [133]. Different from the approaches above, Costa [134] presented a fast deformation model based on the principle of Pascal and the conservation of volume to simulate deformation of soft tissues formed by fibers and fluid. By using the conservation of volume to represent the near incompressibility of soft tissues, the volume displacement in one direction directly causes the displacement of any surface in the opposite direction. This approach is particularly fast, since the variation of liquid pressure within a confined object is constant for the liquid inside the object according to the principle of Pascal and has the same value at all vertices. However, this method is only valid for objects filled with fluids and does not exhibit any dynamic behaviors.

### D. Summary of Representative Deformable Models

Since the real-time computational performance is hardware-dependent and the development of each deformable model is time-varied, Table. I presents a summary of computational performances of representative deformable models in each category, together with their advent times, hardware settings, and surgical applications.

TABLE I
SUMMARY OF REPRESENTATIVE DEFORMABLE MODELS

| Model | Year | Hardware | Speed | Application |
|---|---|---|---|---|
| Deformable spline surface [23] | 1993 | Silicon Graphics (SGI) | 15,000 polygons (real-time) | Laparoscopic gall-bladder surgery |
| FFE [75] | 1998 | 4-processor SGI ONYX | 250 nodes (20 Hz) | Simulation of lower leg |
| 3D ChainMail [44] | 1998 | 8-processor SGI Challenge | 125,000 chain elements (real-time) | Arthroscopic knee surgery |
| Linear TMM [81] | 2000 | Dec Alpha 233 MHz | 4,000 edges (40 Hz) | Hepatic surgery |
| Linear BEM [82] | 2001 | SGI R-4400 | 150 nodes (15 Hz) | Laparoscopic liver surgery |
| Localized linear PCMFS [40] | 2005 | Intel Pentium IV 2.2 GHz | 28 nodes (13.7 ms) | Liver deformation |
| TLED [77] | 2007 | Intel Pentium IV 3.2 GHz | 6,000 hexahedrons (16 ms) | Brain shift in image-guided surgery |
| POD [103] | 2008 | AMD Quad Opteron 2.2 GHz | 8,514 nodes (6 DOFs reduced model) (472-483 Hz) | Palpation of human cornea |
| Data-driven co-rotational FEM [132] | 2009 | Standard PC (specification not available ) | 1,691 tetrahedrons (10 Hz) | Soft tissue deformation |
| MTLED [109] | 2010 | Intel Pentium IV 3.0 GHz | 4,314 nodes (14.9 ms) | Soft tissue deformation |
| Fibers-fluid [134] | 2012 | Intel Pentium 1.66 GHz | 2,000 surface grid points (30 Hz) | Breast deformation |
| PGD [105] | 2013 | Intel Core i7 2.66 GHz | 8,559 nodes (real-time) | Palpation of liver |
| MSM [24] | 2016 | Intel Core i7-2600 3.4 GHz | 15,946 tetrahedrons (33.88 ms) | Laparoscopic gall-bladder surgery |
| PCISPH [119] | 2016 | Intel Core i5-2500k 3.3 GHz | 250,305 particles (481 ms) | Soft tissue deformation |
| Machine Learning [130] | 2017 | Intel Core i7 3.4 GHz | Approx. 394,000 nodes (0.3 s) | Liver deformation |

## III. Linear and Nonlinear Modeling of Soft Tissue Deformation

In this section, the theory of linear elasticity and finite deformation are reviewed, which are commonly used as the physical laws to govern soft tissue behaviors. The physical laws are described by constitutive equations based on continuum mechanics to rigorously define the relationship between stress and strain. The linear elasticity is studied for soft tissue deformation mainly due to its computational efficiency where the stiffness matrix can be pre-computed, inverted and stored. However, to be useful in clinical simulation where higher physical accuracy is required, a method must use fully nonlinear formulations to handle large (finite) deformation of biological tissues [111], provided that the added computational complexity is acceptable.

### A. Linear Elasticity

Due to real-time computational requirement of surgical simulation, most of the existing methods employ the theory of linear elasticity to describe soft tissue deformation [40, 75, 80, 82]. The theory of linear elasticity assumes a linear relationship between the components of stress and strain, and infinitesimal strains. Owing to this linearization, the stress vector $\boldsymbol{\sigma}$ is related to the strain vector $\boldsymbol{\varepsilon}$ by

$$\boldsymbol{\sigma} = \boldsymbol{C}\boldsymbol{\varepsilon} \qquad (1)$$

where $\boldsymbol{C}$ is the tangent matrix of the stress-strain relationship, known as the material matrix, which is constant throughout the simulation; for isotropic and homogeneous materials, this matrix is defined by the two Lamé coefficients $\lambda$ and $\mu$.

Further, the strain components are also linearized based on the assumption of very small deformation gradients and hence neglects the high-order nonlinear terms. The strain vector $\boldsymbol{\varepsilon}$ is related to the displacement $\boldsymbol{u}$ by a shape function $\boldsymbol{B}$

$$\boldsymbol{\varepsilon} = \boldsymbol{B}\boldsymbol{u} \qquad (2)$$

With relations represented in (1) and (2), the strain energy

$W_{strain}$ of a linear elastic model [75] can be written as

$$W_{strain} = \frac{1}{2}\int_{\Omega} \boldsymbol{\varepsilon}^T \boldsymbol{\sigma}\, dx = \frac{1}{2}\int_{\Omega} \boldsymbol{u}^T \boldsymbol{B}^T \boldsymbol{C}\boldsymbol{B}\boldsymbol{u}\, dx \qquad (3)$$

where $\Omega$ is the spatial domain of the linear elastic object, consisting of points at position $\boldsymbol{x}$ with $\boldsymbol{x} \in \Omega$.

After discretization of the spatial domain $\Omega$ into a number of finite elements $\Omega^e$, the strain energy $W_{strain}$ of the object can be represented by the sum of strain energy $W_{strain}^e$ of each individual finite element

$$W_{strain} = \sum_e W_{strain}^e = \frac{1}{2}\sum_e \int_{\Omega^e} \boldsymbol{u}^{eT} \boldsymbol{B}^{eT} \boldsymbol{C}\boldsymbol{B}^e \boldsymbol{u}^e\, dx \qquad (4)$$

The equilibrium state is achieved by minimizing the total potential energy of the system with consideration of the applied external force. The equilibrium equation for each individual element can be written as

$$\int_{\Omega^e} \boldsymbol{B}^{eT} \boldsymbol{C}\boldsymbol{B}^e \boldsymbol{u}^e\, dx = \boldsymbol{f}^e \qquad (5)$$

where $\boldsymbol{f}^e$ is the discretized force vector for element $\Omega^e$.

Since everything inside the integral sign is constant, equation (5) can be reduced to a linear matrix equation in the form of

$$\boldsymbol{K}^e \boldsymbol{u}^e = \boldsymbol{f}^e \qquad (6)$$

where $\boldsymbol{K}^e$ is commonly known as the element stiffness matrix which is represented by

$$\boldsymbol{K}^e = \int_{\Omega^e} \boldsymbol{B}^{eT} \boldsymbol{C}\boldsymbol{B}^e\, dx = \boldsymbol{B}^{eT} \boldsymbol{C}\boldsymbol{B}^e V^e \qquad (7)$$

where $V^e$ is the volume of element $\Omega^e$.

By assembling (6) with respect to all the elements in the object $\Omega$, a linear system of equations governing the equilibrium state of the object can be formed

$$\boldsymbol{K}\boldsymbol{u} = \boldsymbol{f} \qquad (8)$$

where $\boldsymbol{K}$ is the global stiffness matrix of size $3n$ x $3n$, $\boldsymbol{u}$ is the global displacement matrix of size $3n$ x $1$, and $\boldsymbol{f}$ is the global force matrix of size $3n$ x $1$, where $n$ is the number of nodes in the system.

The advantage of employing a linear elastic model for interactive soft tissue deformation is that the global stiffness matrix $\boldsymbol{K}$ is constant throughout the simulation, and therefore its value can be pre-computed in an offline phase to facilitate the online interactive simulation. Despite the reduced runtime computation, using linear elasticity as the basic model involves a few assumptions that limit the accuracy of the physical material being modeled. More importantly, its solution is only valid at small deformation and cannot handle large deformation of soft tissues [135]. Modeling large deformation, such as global rotations, in the framework of linear elasticity often results in an unrealistic volume growth of the model [32]. In order to address the large deformation of soft tissues, many models were proposed by considering geometric nonlinearity for large deformation, such as the geometric nonlinear TMM [32, 136], geometric nonlinear PCMFS [5], and co-rotational FEM [3, 71, 137]. However, due to still using a linear material law, these models suffer from the problem that they can only handle geometric nonlinearity rather than material nonlinearity. Schwartz et al. [138] presented a geometric linear, material nonlinear TMM model, in which the two Lamé coefficients are modified to change the local elastic property for achieving nonlinear deformable modeling. However, this modification does not comply with the constitutive laws of material, since the Lamé constants are the inherent attributes of a material, and they are not changeable constants.

### B. Nonlinear Hyperelasticity

In order to handle the large deformation of soft tissues, in which both geometric and material nonlinearities are involved, the deformable models must be compatible with the theory of finite deformation. The total Lagrangian formulation-based models handle this issue by expressing the stress and strain at a point using the second Piola-Kirchhoff stress $S$ and the Green-Saint Venant strain $E$, which measure the variations of the stress and strain with reference to the initial system configuration. The Green-Saint Venant strain $E$ can be calculated by

$$E = \frac{1}{2}(C - I) \qquad (9)$$

where $I$ is the identity matrix of the second rank, and $C$ is the right Cauchy-Green deformation tensor given by $C = F^T \cdot F$, where $F$ is the deformation gradient given by

$$F = \frac{\partial x}{\partial X} \qquad (10)$$

where $X$ is the position vector of a particle in the initially undeformed configuration, and $x$ is the position vector of the particle in the deformed configuration.

The strain energy $W_{strain}$ is expressed as a function of the components of the Green-Saint Venant strain $E$; subsequently, the second Piola-Kirchhoff stress $S$ is derived from the strain energy $W_{strain}$, which is expressed by

$$S = \frac{\partial W_{strain}}{\partial E} = 2\frac{\partial W_{strain}}{\partial C} \qquad (11)$$

For an isotropic and homogeneous material, the strain energy $W_{iso}$ is a function of the invariants of the right Cauchy-Green deformation tensor, i.e.

$$W_{iso} = W_{strain}(I_1, I_2, I_3) \qquad (12)$$

where $I_1$, $I_2$ and $I_3$ are the three invariants of the right Cauchy-Green deformation tensor $C$, given by

$$I_1 = \text{tr}(C)$$
$$I_2 = \frac{1}{2}[[\text{tr}(C)]^2 - \text{tr}(C^2)] \qquad (13)$$
$$I_3 = \det(C)$$

where $\text{tr}(\cdot)$ and $\det(\cdot)$ denote the trace and determinant of a matrix. Soft tissue material properties can be realized by specifying various strain energy density functions [139].

Similar to (5), the nodal reaction force for each element can be written as

$$\int_{0\Omega^e} {}_0^t B^{eT} {}_0^t \tilde{S} \, d \, {}^0x = {}^t f^e \qquad (14)$$

where the left superscript represents the current time, and the left subscript represents the time of the reference configuration-0 when the total Lagrangian formulation is used [77, 140]. Various total Lagrangian formulation-based methods, such as the TLED [77], total Lagrangian FEM with implicit integration [99], MTLED [109], MJED [98], and total Lagrangian SPH [120] have been studied to handle the large deformation of soft tissues. The details on nonlinear hyperelasticity can be found in [141, 142].

## IV. COMPARISON OF INTERNAL FORCES

Soft tissues are essentially dynamic systems, and their behaviors are governed by the principles of mathematical physics to react to the applied force in a natural manner. Based on Newton's second law of motion, the equation governing the dynamics of soft tissue deformation can be written as

$$m\ddot{u} = f \qquad (15)$$

where $u$ is the displacement vector of a point whose components represent the displacement in $x$, $y$ and $z$ directions, $\ddot{u}$ is the acceleration vector, $m$ represents the mass density of the point, and $f$ represents the net force applied to the point.

The net force $f$ can be further expressed in terms of the internal nodal force $f_{int}$ and the external applied force $f_{ext}$, hence (15) can be written into a form of

$$m\ddot{u} + f_{int} = f_{ext} \qquad (16)$$

Since the internal nodal force $f_{int}$ not only determines the model accuracy but also attributes to the computational complexity of a deformable model, it is worthy comparing the internal nodal forces $f_{int}$ in different deformation methods. Table. II provides a comparison of internal forces for various deformable models.

TABLE II
COMPARISON OF INTERNAL FORCES AMONG DIFFERENT DEFORMABLE MODELS

| Deformable model | Internal force |
|---|---|
| MSM [24] | $f_{i.int} = \sum_{j \in N(P_i)} k_{ij}(\|P_iP_j\| - l_{ij}^0)\frac{P_iP_j}{\|P_iP_j\|}$ |
| Strain energy-based MSM [86] | $f_{i.int} = \sum_{j \in N(P_i)} \frac{\|W_j - W_i\|}{\|P_iP_j\|}\frac{P_iP_j}{\|P_iP_j\|}$ |
| Linear FEM [75] | $f_{int} = Ku = \int_{\Omega_i} B^T CBu \, dx$ |
| Co-rotational FEM [137] | $f_{int} = RKR^T u$ |
| Linear TMM [81] | $f_{i.int} = [K_{ii}]P_i^0 P_i + \sum_{j \in N(P_i)} [K_{ij}]P_j^0 P_j$ |
| Nonlinear TMM [32, 136] | $f_{i.int} = 2\sum_j B_{pj}u_j + 2\sum_{j,k}(u_k \otimes u_j)C_{jkp}$ $+ \sum_{j,k}(u_j \otimes u_k)C_{pjk} + 4\sum_{j,k,l}D_{jklp}u_iu_ku_j$ |
| TLED [77] | $f_{int} = \int_{0_{\Omega^e}} {}_0^t B^{eT} {}_0^t \tilde{S} \, d \, {}^0x$ |
| MJED [98] | $f_{i.int} = \frac{\partial W_h}{\partial P_i}; \, W_h = V_o \sum_k f^k(J)g^k(I)$ |
| HEML [87] | $f_{i.int} = \frac{\partial W_k(I_k)}{\partial P_i}$ |
| Linear PCMFS [40, 143] | $f_{i.int} = \sum_{j \in N(P_i)} K_{ij}\alpha_j = \int_{\Omega_i} B_i^T CB_j\alpha_j \, dx$ |
| MTLED [109] | $f_{i.int} = \sum_{I=1}^{N_{IP}} ({}_0^tF_0^t B^T {}_0^t \tilde{S})|_{x = x_I} W_I$ |
| Normalized total Lagrangian SPH [120] | $f_{i.int} = \sum_{j \in S} V_iV_j(P_j\tilde{\nabla}_X W(R_j, h) - P_j\tilde{\nabla}_X W(R_i, h))$ |
| Fibers-fluid technique [134] | $f_{int} = f_{liquid} + f_{fibres}$<br><br>$= PS_i + \sum_{j \in N(P_i)} \dfrac{-\gamma_{ij}(S_i + S_j)(u_i - u_j)}{2(N-1)\|P_j P_i\|}$ |

MSM: $N(P_i)$ is the set of mass points $j$ at position $P_j$ in the neighborhood of point $i$ at position $P_i$, $k_{ij}$ is the stiffness of the spring connecting points $i$ and $j$, $l_{ij}^0$ is the rest length of spring, and "$\|\cdot\|$" represents the modulus of position vector between two points.

Strain energy-based MSM: $W_i$ is the strain energy at mass point $i$.

Linear FEM: $K$ is the stiffness matrix, $u$ is the displacement, $B$ is the shape function, $C$ is the material matrix, and $x$ is the position of points.

Co-rotational FEM: $R$ is the element rotation matrix of the element local frame with respect to its initial orientation, being updated at each time step.

Linear TMM: $[K_{ii}]$ is the sum of stiffness tensors associated with the tetrahedrons adjacent to $P_i$, $[K_{ij}]$ is the sum of stiffness tensors associated with the tetrahedrons adjacent to edge $(i, j)$, and $P_i^0$ and $P_j^0$ denote the initial positions of points $i$ and $j$.

Nonlinear TMM: $B_{pj}$, $C_{jkp}$ and $D_{jklp}$ are called the stiffness parameters.

TLED: $\tilde{S}$ is the vector form of the second Piola-Kirchhoff stress $S$.

MJED: $W_h$ is the strain energy which is decomposed into $f(J)$ and $g(\bar{I})$, where $\bar{I} = (I_1, I_2 \dots)$ are the invariants of the right Cauchy-Green deformation tensor $C$; hence, $g$ is independent of Jacobian $J$ and its derivative does not involve matrix inversion.

HEML: The strain energy $W_k$ is a function of the squared edge length vector $l_k$.

Linear PCMFS: $\alpha_j$ is the vector of nodal unknown displacement.

MTLED: $N_{ip}$ is the total number of integration points (IP) distributed in the problem domain, ${}^t_0 F$ is the deformation gradient between the undeformed, initial configuration and the configuration at time $t$, and $W_I$ is the weight corresponding to integration point $x_I$.

Normalized total Lagrangian SPH: $V_i$ is the reference volume associated with particle $i$, $P_i$ is the first Piola-Kirchhoff stress tensor at particle $i$, $\tilde{\nabla}_i$ is the gradient operator in the total Lagrangian formulation, $W$ is referred to as the kernel, $R_j$ is the reference distance vector $R_j = X_j - X_i$, where $X_i$ is the position of particle $i$ in the initial configuration, and $h$ is the smoothing length.

Fibers-fluid technique: $P$ is the pressure, $S_i$ is the area of point $i$, $\gamma_{ij}$ is the force per unit of area, $N$ is the number of elastic fibers, and $u_i$ is the displacement vector of point $i$.

## V. Anisotropy, Viscoelasticity, and Compressibility

Biological soft tissues exhibit complex heterogeneous anisotropic, nonlinear, and time- and rate-dependent mechanical behaviors, and they are nearly incompressible [34, 144]. The anisotropic mechanical behaviors arise from the presence of a highly-organized microstructure such as those of connective tissues. These are predominantly composed of collagen or elastin fibers embedded in an amorphous matrix [145] and can be considered as fiber reinforced composites in some cases. Further, the presence of vasculature and other functional components also leads to soft tissue directional dependent behaviors [93]. Regarding viscoelasticity, experimental results have suggested that traditional elastic models provide only a rough approximation of the actual soft tissue response, mainly due to the time-dependent tissue response such as stress relaxation, creep and hysteresis. Collectively, they are called the features of viscoelasticity, leading to the need for viscoelastic models [145]. Finally, most soft biological tissues, such as the brain, liver, kidney and prostate, are usually considered as nearly incompressible [146]. As such, it is necessary for deformable models in surgical simulation to consider these complex biomechanical behaviors in order to achieve realistic simulation of soft tissue deformation.

### A. Anisotropy

Soft tissue anisotropic behaviors can be modeled via a number of techniques in the deformable models mentioned above. The heuristic modeling methodology defines anisotropic materials by assigning different values to the parameters of individual components in the model, such as the stiffness constant of springs in MSM, the geometric limit of chain elements in ChainMail algorithm, and the stiffness of clusters in shape matching technique. Although the anisotropy can be achieved to some extent, these heuristic approaches can only produce arbitrary anisotropic behaviors due to the underlying discrete representation of the model. In contrast, the continuum-mechanical methodology, such as FEM and meshless methods, handles anisotropic materials based on the theory of nonlinear anisotropic elasticity, in which the directional-dependent behaviors of an anisotropic material is accommodated via modification of the strain energy density function.

The strain energy density function is modified for fiber reinforced anisotropic materials by employing unit vectors in the initial configuration to describe local fiber directions [147]. For transversely isotropic materials characterized by fibers dispersion around a preferred fiber direction $a^0$ in the initial configuration, the strain energy density $W_{aniso}$ can be expressed by

$$W_{aniso} = W_{strain}(I_1, I_2, I_3, I_4, I_5) \tag{17}$$

where $I_1$, $I_2$ and $I_3$ are the three invariants of the right Cauchy-Green deformation tensor $C$; and $I_4$ and $I_5$ are the two pseudo-invariants, which arise from the anisotropy introduced by the local fiber $a^0$. $I_4$ and $I_5$ along with their derivatives with respect to $C$ [147] are expressed by

$$I_4 = a^0 \cdot C \cdot a^0; \frac{\partial I_4}{\partial C} = a^0 \otimes a^0$$

$$I_5 = a^0 \cdot C^2 \cdot a^0; \frac{\partial I_5}{\partial C} = a^0 \otimes C \cdot a^0 + a^0 \cdot C \otimes a^0 \tag{18}$$

where $\otimes$ represents the tensor outer product.

The anisotropy stress introduced by the local fiber $a^0$ via $I_4$ and $I_5$ is reflected in the second Piola-Kirchhoff stress $S$ by the additional terms contributed by $I_4$ and $I_5$, i.e.

$$S = 2\left[ \frac{\partial W_{iso}}{\partial C} + \left( \frac{\partial W_{strain}}{\partial I_4}(a^0 \otimes a^0) \right. \right.$$
$$+ \frac{\partial W_{strain}}{\partial I_5}(a^0 \otimes C \cdot a^0 + a^0 \cdot C \tag{19}$$
$$\left. \left. \otimes a^0) \right) \right]$$

For orthotropic materials, which are characterized by three mutually orthogonal preferred directions, unit vectors $a^0$ and $b^0$ are employed to indicate the preferred fiber directions in the initial configuration [93]. The third unit vector orthogonal to the other two vectors is naturally emerged as a preferred direction. The strain energy density $W_{aniso}$ for orthotropic materials can be expressed by

$$W_{aniso} = W_{strain}(I_1, I_2, I_3, I_4, I_5, I_6, I_7) \tag{20}$$

where $I_6$ and $I_7$ are the two pseudo-invariants, which arise from the anisotropy introduced by the local fiber $b^0$. $I_6$ and $I_7$ along with their derivatives with respect to $C$ are expressed by

$$I_6 = \boldsymbol{b}^0 \cdot \boldsymbol{C} \cdot \boldsymbol{b}^0; \quad \frac{\partial I_6}{\partial \boldsymbol{C}} = \boldsymbol{b}^0 \otimes \boldsymbol{b}^0$$

$$I_7 = \boldsymbol{b}^0 \cdot \boldsymbol{C}^2 \cdot \boldsymbol{b}^0; \quad \frac{\partial I_7}{\partial \boldsymbol{C}} = \boldsymbol{b}^0 \otimes \boldsymbol{C} \cdot \boldsymbol{b}^0 + \boldsymbol{b}^0 \cdot \boldsymbol{C} \otimes \boldsymbol{b}^0 \tag{21}$$

The second Piola-Kirchhoff stress $\boldsymbol{S}$ is now given by

$$\boldsymbol{S} = 2 \left[ \frac{\partial W_{iso}}{\partial \boldsymbol{C}} + \left( \frac{\partial W_{strain}}{\partial I_4} (\boldsymbol{a}^0 \otimes \boldsymbol{a}^0) \right. \right.$$
$$+ \frac{\partial W_{strain}}{\partial I_5} (\boldsymbol{a}^0 \otimes \boldsymbol{C} \cdot \boldsymbol{a}^0 + \boldsymbol{a}^0 \cdot \boldsymbol{C}$$
$$\otimes \boldsymbol{a}^0) + \frac{\partial W_{strain}}{\partial I_6} (\boldsymbol{b}^0 \otimes \boldsymbol{b}^0)$$
$$+ \frac{\partial W_{strain}}{\partial I_7} (\boldsymbol{b}^0 \otimes \boldsymbol{C} \cdot \boldsymbol{b}^0 + \boldsymbol{b}^0 \cdot \boldsymbol{C}$$
$$\left. \left. \otimes \boldsymbol{b}^0) \right) \right] \tag{22}$$

### B. Viscoelasticity

Biological soft tissues exhibit time-dependent viscous and elastic characteristics undergoing deformation, which are often referred to as viscoelastic behaviors [145]. Various forms of viscoelastic models were developed in the literature for computational biomechanics, including the Maxwell model [148], Kelvin-Voigt model [149], standard linear solid (SLS) model [150], generalized Wiechert model [151], and double Maxwell-arm Wiechert (DMW) representative model [152]. These models all consist of two types of elemental components: a linear dashpot and a linear spring [152]. In continuum-mechanical methodology, the viscoelastic characteristics of soft tissues are accommodated by using a time-dependent strain energy density function $\widehat{W}_{strain}$. The time-dependent strain energy density function is expressed in the form of a convolution integral [93] as follows

$$\widehat{W}_{strain} = \int_0^t \alpha(t - t') \frac{\partial W_{strain}}{\partial t'} \, dt' \tag{23}$$

where $t$ is time, $t'$ is an arbitrary past time between 0 and $t$, and $\alpha(t)$ is a relaxation function expressed in terms of a Prony series, i.e., $\alpha(t) = \alpha_\infty + \sum_{i=1}^N \alpha_i e^{-t/\tau_i}$ with positive constants $\alpha_\infty$, $\alpha_i$ and $\tau_i$. Such form for the relaxation function is a generalized Maxwell model [148]. By imposing condition $\alpha_\infty + \sum_{i=1}^N \alpha_i = 1$, $\alpha(t)$ can be further written as

$$\alpha(t) = 1 - \sum_{i=1}^N \alpha_i \left( 1 - e^{-t/\tau_i} \right) \tag{24}$$

The corresponding second Piola-Kirchhoff stress $\boldsymbol{S}$ is modified to a time-dependent stress tensor $\widehat{\boldsymbol{S}}$ by considering the time-dependent strain energy density function $\widehat{W}_{strain}$, i.e.

$$\widehat{\boldsymbol{S}} = 2 \frac{\partial \widehat{W}_{strain}}{\partial \boldsymbol{C}} = \int_0^t \alpha(t - t') \frac{\partial \boldsymbol{S}}{\partial t'} \, dt'$$
$$= \int_0^t \left[ 1 - \sum_{i=1}^N \alpha_i \left( 1 - e^{(t'-t)/\tau_i} \right) \right] \frac{\partial \boldsymbol{S}}{\partial t'} \, dt' = \boldsymbol{S} - \sum_{i=1}^N \boldsymbol{\gamma}_i \tag{25}$$

where $\boldsymbol{\gamma}_i$ is a time-dependent term given by

$$\boldsymbol{\gamma}_i = \int_0^t \alpha_i \left( 1 - e^{(t'-t)/\tau_i} \right) \frac{\partial \boldsymbol{S}}{\partial t'} \, dt' \tag{26}$$

Equation (26) can be converted into an incremental update applied at each time increment $\Delta t$ after a discretization over time

$$\boldsymbol{\gamma}_i{}^t = A_i \boldsymbol{S}^t + B_i \boldsymbol{\gamma}_i{}^{t-\Delta t} \tag{27}$$

where constant coefficients $A_i$ and $B_i$ are determined by

$$A_i = \frac{\Delta t \alpha_i}{\Delta t + \tau_i}; \quad B_i = \frac{\tau_i}{\Delta t + \tau_i} \tag{28}$$

### C. Compressibility

When there is a certain compressibility involved in the deformation of isotropic soft tissue elastic models, the strain energy density function may be decomposed into a volumetric part $W_{strain}^{vol}$ and an isochoric part $W_{strain}^{iso}$ [94], which separates the isochoric (volume-preserving) and volumetric components. The strain energy density function can be written as

$$W_{strain} = W_{strain}^{iso}(\overline{I_1}, \overline{I_2}) + W_{strain}^{vol}(J) \tag{29}$$

where $\overline{I_1}$ and $\overline{I_2}$ are the invariants of the isochoric part of the right Cauchy-Green deformation tensor, and $J$ is the volume ratio. $\overline{I_1}$, $\overline{I_2}$ and $J$ can be calculated by

$$\overline{I_1} = \mathrm{tr}(\overline{\boldsymbol{C}}); \quad \overline{I_2} = \frac{1}{2} [[\mathrm{tr}(\overline{\boldsymbol{C}})]^2 - \mathrm{tr}(\overline{\boldsymbol{C}}^2)]$$
$$J = \det(\boldsymbol{F}) \tag{30}$$

where $\overline{\boldsymbol{C}}$ is the isochoric part of the right Cauchy-Green deformation tensor given by $\overline{\boldsymbol{C}} = J^{-2/3} \boldsymbol{C}$.

## VI. NUMERICAL TIME INTEGRATION

Due to dynamic process of soft tissue deformation, it is necessary to integrate the dynamic equation in the temporal domain for soft tissue modeling in surgical simulation. Currently, the dynamics of soft tissue deformation are commonly obtained by numerical time integration schemes such as the explicit [81, 87] and implicit [3, 153] integrations. In both schemes, the second-order ordinary differential equation governing the dynamics of soft tissue deformation is transformed into a coupled set of two first-order systems by introducing a proxy velocity vector $\dot{\boldsymbol{u}}$, i.e.

$$\begin{cases} \ddot{\boldsymbol{u}} = \dfrac{\dot{\boldsymbol{u}}}{t} \\ \dot{\boldsymbol{u}} = \dfrac{\boldsymbol{u}}{t} \end{cases} \tag{31}$$

where $\ddot{\boldsymbol{u}}$, $\dot{\boldsymbol{u}}$ and $\boldsymbol{u}$ are the acceleration, velocity and displacement vectors at a point.

To determine a solution to the dynamic equation (31), the time-dependent variables $\dot{\boldsymbol{u}}$ and $\ddot{\boldsymbol{u}}$ are discretized using a finite difference technique via a time increment to estimate the continuous variables [154]. By choosing different finite difference techniques, such as the forward or backward finite difference estimates, an explicit or implicit integration scheme can be obtained [155].

### A. Explicit Integration

In the explicit integration, variables in the future state are explicitly determined from their current state of known values via numerical time-stepping. An explicit scheme using the forward finite difference estimation can be written as

$$\begin{cases} \ddot{\boldsymbol{u}}^t = \dfrac{\dot{\boldsymbol{u}}^{t+\Delta t} - \dot{\boldsymbol{u}}^t}{\Delta t} \\ \dot{\boldsymbol{u}}^t = \dfrac{\boldsymbol{u}^{t+\Delta t} - \boldsymbol{u}^t}{\Delta t} \end{cases} \rightarrow \begin{cases} \dot{\boldsymbol{u}}^{t+\Delta t} = \dot{\boldsymbol{u}}^t + \Delta t \ddot{\boldsymbol{u}}^t \\ \boldsymbol{u}^{t+\Delta t} = \boldsymbol{u}^t + \Delta t \dot{\boldsymbol{u}}^t \end{cases} \tag{32}$$

where the right superscript denotes the current and future time points at $t$ and $t + \Delta t$, respectively, and $\Delta t$ is the time step.

The explicit integration is easy to implement and computationally efficient, since variables in the future state are obtained explicitly based on the current state of known values only, without requiring the inversion of stiffness matrix at each time step [75]. It is also well suitable for distributed parallel computing, since most of the deformable models for soft tissue simulation employ the mass lumping technique, by which the global system of equations can be split into independent equations for individual nodes, allowing each node to be assigned to a processor in the parallel computer to perform calculations independently. Despite the computational efficiency and simple implementation, the explicit integration exhibits a number of shortcomings. More importantly, the solution of the explicit integration is only conditionally stable. It requires a careful selection of the time step for a simulation to be stable, otherwise the simulation will explode numerically [8]. The mathematical evaluation of the stability of an integration scheme can be conducted using the Dahlquist's test equation [156]

$$\dot{y} = \lambda y(t), \quad y(t_0) = y_0 \qquad (33)$$

with the analytic solution given by $y(t) = y_0 e^{\lambda t}$, where $\lambda$ is a constant.

An integration scheme that yields a bounded solution to (33) is said to be stable. Equation (33) is only bounded when $\Re e(\lambda) \leq 0$. Using the explicit integration, equation (33) can be approximated as

$$\dot{y}^n = \frac{y^{n+1} - y^n}{\Delta t} = \lambda y^n \qquad (34)$$

Equation. (34) can be further arranged into

$$y^{n+1} = \lambda \Delta t y^n + y^n = (1 + \lambda \Delta t)^{n+1} y_0 \qquad (35)$$

where the condition for $y^{n+1}$ not to increase indefinitely is

$$|1 + \lambda \Delta t| \leq 1 \qquad (36)$$

It can be seen from (36) that the explicit integration is only conditionally stable, and the critical time step $\Delta t$ is obtained by $\Delta t \leq \frac{2}{|\lambda|}$. In dynamic soft tissue deformation, this means that the time step must meet the Courant-Friedrichs-Lewy (CFL) condition for numerical stability [157], i.e., the time step must not exceed the time for a stress wave to traverse the smallest element in the finite element mesh. Mathematically, the maximum time step is associated with the largest eigenvalue of the stiffness matrix and the mass and damping values [81, 140]. Various estimations of the critical time step for stable simulation of soft tissues in the explicit integration were studied, such as the critical time step for linear FEM [158], TLED [77] and meshless method [110]. Due to the stiff equations raised from the near incompressibility of biological soft tissues, the maximum time step is often restricted to be a small value. Further, the soft tissue viscoelastic effect further decreases the maximum value of the time step. Owing to the small time step, the solutions of the explicit integration to soft tissue deformation usually require more iterations per simulation frame, resulting in inefficient computation.

To address the problem of inefficient computation resulted from using a small time step in the explicit integration, various techniques were studied. Cotin et al. [81] applied a fourth-order Runge-Kutta explicit integration scheme to discretize the temporal domain and achieved a larger time step, which is about 10x larger than that of the forward Euler method, leading to a speed gain of 2x for surgical simulation. Fierz et al. [154] studied a shape matching technique to increase the time step size in the explicit integration. In this approach, under the given desired simulation time step, the ill-shaped elements that cannot be simulated stably by a standard deformation model are handled specially via a non-physics-based geometric shape matching technique. The elements that require the special treatment are identified by computing the eigenmodes of the elements while considering the mutual interactions with their neighboring elements. This approach enables taking a larger time step than the standalone explicit integration, and the total computational costs per frame are significantly reduced. Taylor et al. [76] presented a reduced order explicit dynamics scheme to improve the time step limit. In this approach, the full model configuration is projected onto a lower dimensional generalized basis prior to the integration of the equilibrium equation, and hence the time integration is performed on a reduced basis, leading to a much larger time step than that on the full system for stable simulation. However, it should be noted that the conditional stability of the explicit integration cannot be completely eliminated by all the efforts mentioned above.

### B. Implicit Integration

Compared to the explicit integration, the implicit integration is unconditionally stable. In the implicit integration, variables in the future state are determined by considering variables both in the current and future states, leading to a system of equations in which the unknown state variable values are implicitly given as solutions. An implicit scheme using the backward finite difference estimation can be written as

$$\begin{cases} \ddot{u}^t = \dfrac{\dot{u}^t - \dot{u}^{t-\Delta t}}{\Delta t} \\ \dot{u}^t = \dfrac{u^t - u^{t-\Delta t}}{\Delta t} \end{cases} \qquad (37)$$

If writing the index of $u$ by $t \to t + \Delta t$ and $t - \Delta t \to t$, the time-continuous variables $\ddot{u}$, $\dot{u}$ and $u$ can be estimated as

$$\begin{cases} \dot{u}^{t+\Delta t} = \dot{u}^t + \Delta t \ddot{u}^{t+\Delta t} \\ u^{t+\Delta t} = u^t + \Delta t \dot{u}^{t+\Delta t} \end{cases} \qquad (38)$$

Similar to the stability verification using the Dahlquist's test equation for the explicit integration, equation (33) can be approximated using the implicit integration as

$$y^{n+1} = \lambda \Delta t y^{n+1} + y^n$$
$$\to y^{n+1} = \frac{y^n}{(1 - \lambda \Delta t)} = \frac{y_0}{(1 - \lambda \Delta t)^{n+1}} \qquad (39)$$

where $y^{n+1}$ will not be increased indefinitely if

$$|1 - \lambda \Delta t| \geq 1 \qquad (40)$$

Equation (40) is always true since $\Re e(\lambda) \leq 0$; therefore, the implicit integration is unconditionally stable for any arbitrarily chosen time step [156]. This attribute provides a unique strength to the implicit integration in handling collisions occurred in tool-tissue interactions and the stiff equations raised from the near incompressibility of soft tissues, facilitating the simulation by using a large time step without loss of numerical stability. The time step size is limited only by the factors of numerical convergence and accuracy. Despite its unconditional stability, the implicit integration is computationally more expensive than the explicit counterpart. It requires a solution of a nonlinear system of equations at each load step, which is usually solved by an iterative method based on the Newton-Raphson method through a sequence of solutions of linear equations. The linear system of equations can either be solved by directly computing the inverse or the factorization of the system matrix, or iteratively solving a system of algebraic equations based on an initial estimate, both leading to an increase in computational time. As demonstrated in [81], the computation at one iteration step by the implicit integration is at least one order of magnitude larger than that by the explicit integration. Mafi and Sirouspour [99] studied a method of element-by-element preconditioned conjugate gradients (PCG) in comparison with the conventional PCG for solving the system equations. It is shown that the element-by-element PCG outperforms the conventional PCG at a small number of iterations. In the implicit integration, iterations also need to be performed at each time step in order to control numerical errors and to avoid numerical divergence [77]. Further, since numerical dissipation becomes dominant at a large time step, the solution accuracy will be deteriorated when a large time step is used in low-order schemes [154].

In comparison between both explicit and implicit integration schemes, the system response is more global within the implicit approach, whereas the response of the explicit integration to the applied force is only propagated from a node to the whole mesh after multiple iterations [75]. Due to avoidance in solving a large system of equations, the explicit integration is computationally efficient in finding solutions over the implicit integration. However, the time discretization error is accumulated in the explicit scheme [90]. Further, in the case of large deformation of soft tissues, elements will become distorted and ill-conditioned, leading to a decrease in the critical time step size for the explicit integration [72], whereas the implicit integration still remains stable. The explicit integration also tends to converge more slowly than the implicit counterpart, since it is only conditionally stable. Joldes et al. [159] facilitated the convergent rate of the explicit integration using the concept of dynamic relaxation (DR), and it was further improved by the adaptive DR [160]. The DR is able to increase the convergence rate towards the final deformed state by including a mass proportional numerical damping but at the cost of sacrificing numerical accuracy. The DR-based explicit integration is computationally efficient in that the main DR parameters can be pre-computed. The implicit integration also helps to obtain robust and realistic behaviors when simulating tool-tissue interactions, whereas the explicit integration with a much lower time step and large frame rate results in very damped motions [98]. Moreover, the explicit integration does not guarantee that, at each time step, the residual vector is minimized, and hence it cannot ensure that the external and internal forces are balanced [3], whereas the accuracy of the equilibrium equation can be controlled in the implicit integration [161].

## VII. DISCUSSION

Among various deformable models proposed for modeling of soft tissue deformation, it is obvious that there is no single deformable model that can address both requirements of realistic and real-time surgical simulation. Instead, they were developed in different ways to meet specific needs. Despite their computational advantage, the geometrically-based models are seldom or no longer used for soft tissue deformation due to their non-physics-based nature. MSM is often used when computational efficiency is preferred to the physical accuracy, and it has been used in many commercially available surgical simulators [162, 163]. However, a more accurate model is needed in order to be compatible with clinical practice, provided that the added computational complexity is acceptable. In addition, the optimization process in MSM is a time-consuming task and may lead to failure if prior assumed conditions are changed during the simulation. The ChainMail algorithm is well suitable for modeling of interactive deformation of large medical volumes, where other deformable models cannot achieve the real-time computational performance. The shape matching approach can be used for stable simulation of soft tissue deformation, due to its unconditional stability offered by being a geometric approach [66]. It should be noted that these deformable models exhibit difficulties in determining model parameters to be associated with the constitutive laws governing the mechanical behaviors of soft tissues, and hence they can only produce a physically plausible simulation.

Higher model accuracy can be achieved by the continuum-mechanical methodology despite the increased computational complexity. FEM is often employed for computation of soft tissue deformation if physical accuracy is concerned. Most simplifications made to FEM to facilitate its computational performance inevitably compromise its model accuracy and limit its capability in handling soft tissue material properties. Although real-time computational performance can be achieved with the total Lagrangian-based FEMs, they are only suitable for modeling of soft tissue response that does not involve topology changes induced by surgical operations such as cutting and tearing. The result of pre-computation at the initial system configuration would become invalid when a topology-changing cut is introduced to the system. Model reduction is a promising technique for real-time simulation of soft tissue deformation; however, most of the models developed using this scheme involve an offline and online computation, posing challenges for simulation of topology changes. Despite the recent progress [164] in the framework of model reduction where topology changes are handled by an extended FEM (X-FEM) for the incorporation of discontinuities in the displacement field, the offline computation assumes a certain model behavior, leading to inaccuracy if the model is changed during the online simulation. In addition, it is also important to consider the element-related issues when using the mesh-based approach due to the use of low-order finite element formulations. The hourglass control algorithm and the locking-free tetrahedrons provide the means to mitigate numerical inaccuracies. When simulating large deformation and discontinuities in soft tissue deformation, the meshless approach is often preferred to the mesh-based approach, since it can conduct deformation without explicit construction of nodal connectivity, avoiding most of the element-related issues, such as element distortion and element inversion. Meanwhile, it also avoids the process of mesh generation, facilitating clinical integration of the computer-assisted surgery [21]. However, the accuracy of the meshless approach is heavily dependent on the distribution of particles in the problem domain, and it is only accurate in terms of the global reaction force other than the local reaction force.

The emerging of the neural network approach and machine learning also takes an important role in realistic and real-time surgical simulation. The continuum-mechanical methodology considers a soft tissue as a continuum medium whose behaviors are governed by constitutive laws expressed by partial differential equations. Despite high level of physical accuracy achieved by the continuum-mechanical methods, it is arguable that soft tissues are more complex than idealized continuum models, in terms of both material composition and structure formation. The machine learning technique, on the other hand, seeks for a mapping function through training of a supervised neural network to generate a desired output (soft tissue global deformation) for future unseen inputs (prescribed displacements). The soft tissue mechanical behaviors are encoded implicitly in the trained coefficients of the neural network. Owing to the recent advancement in artificial intelligence and open-source software package such as the TensorFlow [165], many different soft tissues can be employed for training of the neural networks. However, it needs to be noted that the simulated results are highly dependent on the learning algorithm chosen.

TABLE III
COMPARISON OF MAIN DEFORMABLE MODELS

| Deformable Model | Accuracy | Speed | Remark |
|---|---|---|---|
| Geometrically-based | * | **** | Lack of deformable physics |
| MSM | *** | **** | Generic simulation, such as surgical training |
| ChainMail | * | ***** | Large medical volumes |
| FEM | ***** | *** | Good for scientific analysis |
| Meshless method | **** | *** | Large deformation and discontinuities |
| Machine learning | **** | **** | Rely on training samples and learning algorithms |
| Data-driven | **** | **** | Require patient-specific data |

Scheme: * is the lowest whereas ***** is the highest.

Given the wide varieties and variations of deformable models for surgical simulation, Table. III summaries the capabilities of the main deformation approaches in terms of physical accuracy and computational performance. It can be used as reference to make up an appropriate deformation strategy according to different surgical simulation conditions.

Topology changes due to surgical operations such as cutting and tearing is a requirement in surgical simulation which further complicates deformable modeling. These operations are difficult to achieve within the constraint of real-time performance as topology changes, often involve mesh and surface reconstruction, need to be calculated and updated. In the heuristic modeling methodologies such as MSM and ChainMail, the topology changes are generally accommodated by removing the springs or chain links that are encountered along the path of the cutting tool as the surgical tool passes through the soft tissues [45]. In continuum-mechanical methodology, the topology changes can be accommodated by removing or splitting finite elements, employing X-FEM technique, or modeling discontinuity in the meshless method. These methods for handling topology changes usually require update of the object stiffness matrix and may degrade the critical time step enforced by the explicit time integration, posing numerical challenges for real-time computational performance. The details on physically-based simulation of cutting in deformable models can be found in [166].

To verify the physical accuracy of deformable models for surgical simulation, most of the existing models compared their deformation solutions with those of the FEM reference solutions, such as the solution of commercially available ABAQUS [167] using implicit solver with hybrid formulation of linear elements [109]. However, given that finite element modeling is an approximation method in itself, the accuracy of its results heavily relies upon the quality of its input [6]. Kerdok et al. [6] presented a Truth Cube which sets the practical physical standards for validation of real-time soft tissue deformation models. A cube of silicone rubber with a pattern of embedded Teflon spheres is undergone uniaxial compression and spherical indentation tests, and the cube is scanned by a CT scanner (see Fig. 9). The volumetric displacement results, along with details of the cube construction and boundary conditions in the two loading tests serve as the physical standard for model validation. Despite the available data for validation, it needs to be noted that the Truth Cube has a regular model geometry and well-characterized material properties and loading conditions, but the surgical simulation needs to handle more challenging conditions, involving large deformation, irregular shapes, and complex materials.

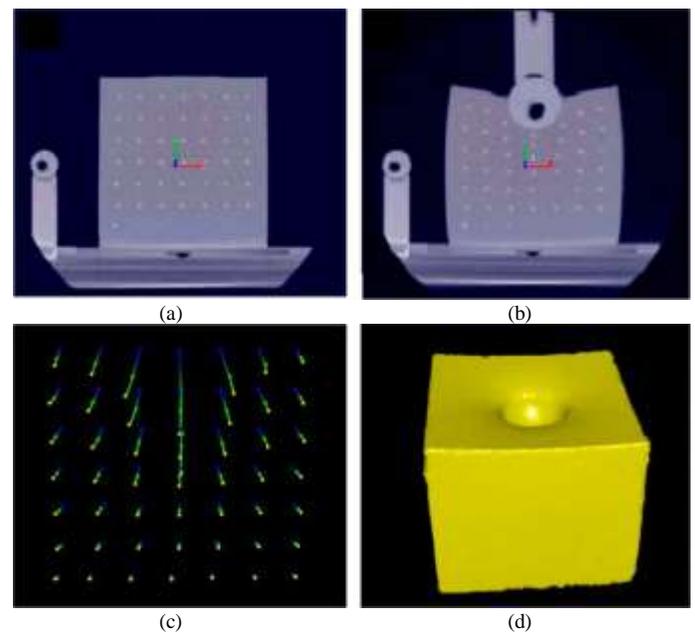

(a)          (b)

(c)          (d)

Fig. 9. (a) A CT scan of a center vertical slice for spherical indentation in undeformed, initial configuration; (b) deformation of the cube under 30% nominal strain; (c) the trajectory and locations of the internal spheres where blue represents no indentation, green represents 22% nominal strain case, and yellow represents 30% nominal strain case; and (d) the surface for the 30% strain case [6].

To be compatible with clinical practice, a deformable model must use patient-specific tissue properties; however, they are significantly difficult to determine for human soft tissues. The first reason is the evident difficulty in carrying out quantitative empirical measurements of human tissues (such as liver) *in vivo*. Second, there are always uncertainties in patient-specific properties of tissues since the mechanical properties of soft tissues obtained through *in vivo* and *in situ* measurements are different from those obtained through *in vitro* measurements [144]. However, despite these reasons, it is still possible to determine deformation of soft tissues during surgery without the knowledge of patient-specific properties of tissues. As evident in [146], the computational biomechanics problems can be reformulated in such a way that the results are weakly sensitive to the variation in mechanical properties of simulated tissues. In particular, the problems can be formulated into (I) pure-displacement and displacement zero traction problems whose solutions in displacement are weakly sensitive to the mechanical properties of the considered continuum; and (II) problems that are approximately statically determinate and thus their solutions in stresses are weakly sensitive to the mechanical properties of constituents. It is shown that good results can be expected for the brain tissues while using even the simplest constitutive model without the knowledge of patient-specific properties [146]. However, it should be noted that the achieved accuracy is limited due to the lack of patient-specific properties.

In general, deformable models play a fundamental role in the development of surgical simulation and will have a wide impact on the development of computer integrated surgery (CIS) system in the near future [1]. Currently, simulation software suites, such as the Simulation Open Framework Architecture (SOFA) [168, 169], finite elements for biomechanics (FEBio) [170, 171], and open-source finite element toolkit (NiftySim) [94], have enabled a wide range of medical applications (see Fig. 10), such as the interactive training system for interventional electrocardiology procedures [74], preoperative trajectory planning for percutaneous procedures [172], modeling of biomechanics of human liver during breathing [130], and biomechanically guided prone-to-supine image registration of breast magnetic resonance images (MRI) [173]. The development of deformable models have also advanced the development of many medical applications such as the tele-surgery for robotic surgery training [174], surgical and interventional robotics [175], Chinese acupuncture training system [176], modeling of needle insertion [177], computer-assisted interventions [178, 179], and myringotomy simulation [180]. Further, the benefits of soft tissue modeling are useful not only for training, planning and control of surgical procedures, but also for optimizing surgical tool design, creating "smart" instruments capable of assessing pathology or force-limiting novice surgeons, and understanding tissue injury mechanisms and damage thresholds [7, 181]. Despite recent progress in deformable modeling, the issues on error control [161] and clinical validation [9] still remain largely open topics, which could further facilitate the integration of CIS and surgical robotics into clinical use, leading to great benefits in medicine.

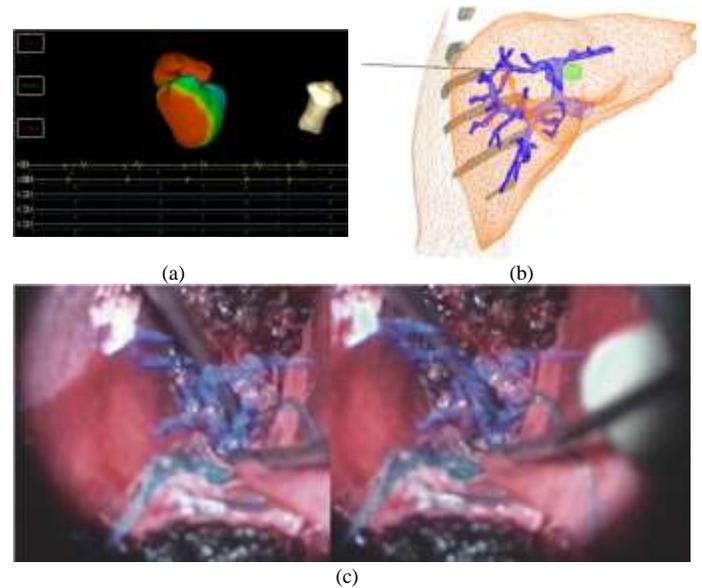

Fig. 10. (a) An interactive training system for interventional electrocardiology procedures [74]; (b) a simulation scene of trajectory planning for percutaneous procedures [172]; and (c) a stereo pair showing MRI derived vasculature beneath the visible resected tissue surface [179].

## VIII. CONCLUSIONS

This paper presents the state-of-the-art soft tissue deformable modeling for interactive surgical simulation. Owing to the realistic and real-time challenges of surgical simulation, various deformable models were studied in the literature to address these issues. The paper classifies the existing deformable models into three main categories, each discussed in detail. Subsequently, it discusses linear and nonlinear deformable modeling, model internal forces, numerical time integration schemes, and modeling of complex biomechanical behaviors such as anisotropy, viscoelasticity, and compressibility. Various issues related to deformable models, topology changes, model validation, patient-specific properties of tissues, medical applications, and clinical impact are also discussed.

The future research directions for further improvement of soft tissue modeling include the following three aspects:

- Physiological modeling to take into account the functional nature of biological tissues. Currently, the existing methods are mainly dominated by physics-based modeling to describe passive mechanical behaviors of soft tissues. However, biological soft tissues also exhibit electrical activities to generate biological functions. Therefore, it is necessary to extend physical modeling to physiological modeling to describe the interaction between passive mechanical deformation and active electrical activities of soft tissues.
- Integration of tissue property measurement into soft tissue modeling. Currently, most of the research efforts on soft tissue modeling have only focused on deformation analysis for surgical simulation, without consideration of tissue property measurement. These methods assume soft tissue properties are fixed and already known, unable to populate with measurement data from real patient-specific tissues. Therefore, the current soft tissue modeling efforts cannot adapt to clinical surgical practice with patient-specific tissue conditions. Soft tissue modeling based on measurement of patient-specific tissue properties will significantly improve the accuracy of soft tissue deformation, thus enhancing the realism of surgery simulation with consideration of patient-specific conditions.

- Incorporation of soft tissue modeling into robotic control loop for surgical and interventional robotics. Soft tissue deformation behaviors are indispensable for robotic surgical planning and procedures. However, the majority of soft tissue modeling is focused on deformation analysis for surgical simulation, without the connection to surgical robot. The incorporation of soft tissue modeling into robotic control loop to drive surgical robot to carry out surgical operations will enable automatic planning and precise control of robotic surgical tasks. It will seamlessly bridge the gap between surgery simulation and surgical practice. It will also revolutionize the discipline of surgical simulation from pure modeling and analysis of surgical processes to the utilization for steering, guidance and control of practical surgical processes, and thus generating a significant impact to robotic surgery.